\documentclass[twoside]{dis09}
\usepackage[latin1]{inputenc}
\usepackage[dvips]{graphicx,epsfig,color}
\usepackage{wrapfig,rotating}
\usepackage{amssymb,amsmath,array}

\begin{document}
\title{A QCD analysis of ZEUS data including DIS inclusive cross sections with longitudinally polarised leptons and data run at lower proton beam energies.}


\author{A M Cooper-Sarkar
%
\thanks{On behalf of the ZEUS Collaboration~\cite{url}}
%
\vspace{.3cm}\\
%
 University of Oxford - Dept of Physics \\
Denys Wilkinson Bdg, Keble Rd, Oxford, OX1 3RH - UK
%
}

\maketitle

\begin{abstract}
New ZEUS data are added into the NLO QCD analysis of the ZEUS-JETS PDF fit. 
The addition of high-$Q^2$ NC and CC $e^-p$ inclusive cross-section data 
improves the determination of the $u$-valence quark at high $x$. 
The addition of high-$Q^2$ CC $e^+p$ inclusive cross-section data 
improves the determination of the $d$-valence quark at high $x$. The addition 
of lower-$Q^2$ NC $e^+p$ inclusive cross-section data, run at three different 
proton beam energies, improves the determination of the sea and gluon PDFs at 
small $x$. The new PDF fit is called the ZEUS09 PDF fit.
\end{abstract}

\section{Introduction}
\label{sec:intro}
The kinematics
of deep inelastic lepton hadron scattering (DIS) is described in terms of the variables $Q^2$, the negative
invariant mass of the exchanged vector boson, Bjorken $x$, the fraction
of the momentum of the incoming nucleon taken by the struck quark (in the 
quark-parton model), and $y$ which measures the energy transfer between the
lepton and hadron systems.
The double differential 
cross-sections for the neutral current (NC) process with lepton polarization $P$ are given by,

\begin{equation}
{d^2\sigma(e^{\pm}p)\over dxdQ^2}={2\pi\alpha^2\over xQ^4}
\left[ H_0^\pm+PH_P^\pm\right],~~H_{0,P}^\pm = Y_+F_2^{0,P} -y^2 F_L^{0,P} \mp Y_-xF_3^{0,P}
\end{equation}
where, $Y_{\pm}=1 \pm (1-y)^2$, and, at LO in QCD, the structure functions $F_2^{0,P}$ and $xF_3^{0,P}$ are directly related to quark distributions by
\begin{equation}
F_2^{0,P} = \sum_ix(q_i+\bar{q}_i)A_i^{0,P},  
xF_3^{0,P} = \sum_ix(q_i-\bar{q}_i)B_i^{0,P}
\end{equation}
The coefficients $A^0_i,B^0_i$ for unpolarised beams are given by
\begin{equation}
        A_i^0(Q^2) =  e_i^2-2e_iv_iv_eP_Z+(v_e^2+a_e^2)(v_i^2+a_i^2)P_Z^2 
\label{eq:unpol-a}
\end{equation}
\begin{equation}
        B_i^0(Q^2) =  -2e_ia_ia_eP_Z+4a_iv_iv_ea_eP_Z^2
\label{eq:unpol-b}
\end{equation}
The 
coefficients for the polarisation terms are given by
\begin{eqnarray}
A^P_i &=& 2e_ia_ev_iP_Z-2a_ev_e(v_i^2+a_i^2)P_Z^2, 
\label{eq:pol-a} \\
B^P_i &=& 2e_ia_iv_eP_Z-2a_iv_i(v_e^2+a_e^2)P_Z^2.
\label{eq:pol-b}
\end{eqnarray}
The term in $P_Z$ arises from $\gamma Z^0$ interference and the term in
$P_Z^2$ arises purely from $Z^0$ exchange, where $P_Z$ accounts for the effect 
of the $Z^0$ propagator relative to that of the virtual photon, and is given by
\begin{equation}
           P_Z = \frac {Q^2} {Q^2 + M_Z^2} \frac {1} {\sin^2 2\theta_W}.
\end{equation}
The other factors in the expression for $A$ and $B$ are
the charge, $e_i$, NC electroweak vector, $v_i$, and axial-vector, $a_i$, 
couplings of quark $i$ and the corresponding NC electroweak 
couplings of the electron, $v_e,a_e$. 
Equations~\ref{eq:pol-a},\ref{eq:pol-b} show that
 polarization effects are only important at high $Q^2$. 

Unpolarized HERA 
data have been used in fits to determine Parton Distribution Functions (PDFs).
For low $x$, $x \leq 10^{-2}$, $F_2^0$ 
is sea quark dominated and its $Q^2$ evolution, as predicted by QCD, is 
controlled by
the gluon contribution, such that HERA data provide 
crucial information on low-$x$ sea-quark and gluon distributions.
At high $Q^2$, the structure function $xF_3^0$ becomes increasingly 
important, and gives information on valence quark distributions. 
The charged current (CC) interactions also
enable separation of the flavour of the valence distributions 
at high-$x$, since their (LO) cross-sections are given by, 
\[
\frac {d^2\sigma(e^+ p) } {dxdQ^2} = (1+P)\frac {G_F^2 M_W^4} {(Q^2 +M_W^2)^2 2\pi x}
x\left[(\bar{u}+\bar{c}) + (1 - y)^2 (d + s) \right],
\]
\[
\frac {d^2\sigma(e^- p) } {dxdQ^2} = (1-P)\frac {G_F^2 M_W^4} {(Q^2 +M_W^2)^2 2\pi x}
x\left[(u + c) + (1 - y)^2 (\bar{d} + \bar{s}) \right].
\]

Parton Density Function (PDF) determinations are usually global 
fits~\cite{mrst,cteq,zeus-s}, which use fixed target 
DIS data as well as HERA data. In such analyses, the high statistics HERA NC 
$e^+p$ data have determined the low-$x$ sea and 
gluon distributions, whereas the fixed target data have determined 
the valence distributions. Now that high-$Q^2$ HERA data on NC and CC
 $e^+p$ and $e^-p$ inclusive double 
differential cross-sections are available, PDF fits can be made to HERA 
data alone, since the HERA high $Q^2$ cross-section 
data can be used to determine the valence distributions. This has the 
advantage that it eliminates the need for heavy target corrections, which 
must be applied to the $\nu$-Fe and $\mu D$ fixed target data. Furthermore
there is no need to assume isospin symmetry, i.e. that $d$ in the 
proton is the same as $u$ in the neutron, 
since the $d$ distribution can be obtained directly from CC $e^+p$ data. 

The ZEUS-JETS PDF fit~\cite{zeus-j} was an NLO QCD fit in the DGLAP formalism 
to ZEUS inclusive cross-section data and jet production data from HERA-I. 
The PDFs were parametrized at $Q^2_0=7$GeV$^2$ by the form 
$xf(x)=p_1 x^{p_2} (1-x)^{p_3} (1+p_4 x)$, using 11 free parameters. 
Predictions for the cross-sections were made by evolving the PDFs to the 
$Q^2$ values of the measurements and convoluting them with coefficient 
functions, calculated in the general mass variable flavour-number scheme of 
Thorne and Roberts~\cite{rtvfn}, to produce structure function predictions. 
Predictions for jet cross-sections were made 
using NLO programmes~\cite{ridolfi,disent}. 
In evaluating the uncertainty on the PDF parameters and the cross-section 
predictions which derive from them full acount is taken of correlated 
experimental uncertainties using the Offset method.  
However the determinations of the valence PDFs from HERA-I data 
are not as accurate as those from 
global fits because of poor statistics at high-$x$. The addition of data from 
HERA-II changes this situation. In order to assess the impact of this new data 
the fit formalism has not been changed. This paper describes the improvement to
 the valence PDFs from adding ZEUS $e^-p$ NC~\cite{DESY-08-202} from the 2005-6
running period, $e^-p$ CC~\cite{DESY-08-177} data from the 
2004-6 running period, and $e^+p$ CC data~\cite{ZEUS-prel-09-002} 
from the 2006-7 running period, into the ZEUS-JETS fit. 

The polarization of the HERA-II data can also 
be exploited to measure electroweak couplings. The CC cross-sections 
give information on 
the propagator mass and the weak coupling. The NC cross-sections give 
information on the quark couplings to $Z^0$. A preliminary model independent 
extraction of these parameters was given in Ref~\cite{ew08} and will be updated
when the full HERA-II data set is available. In the present paper we compare 
our data to the electro-weak predictions of the Standard Model.   

Most of the data collected at HERA were for collisions between 
$e^{\pm}$ of $27.5$GeV with protons of $920$GeV beam energies. 
During the final running period, HERA provided lower $Q^2$ data 
($24 < Q^2 < 110~$GeV$^2$), with modified trigger conditions, collected
 at three different proton beam energies 
($920, 575, 460$GeV). These data access high-$y$ and have been used to measure 
the longitudinal structure function $F_L$~\cite{DESY-09-046}. 
At low-$x$, NLO QCD 
in the DGLAP formalism predicts that 
this structure function is strongly related to the gluon PDF. 
The reduced cross-section data from these runs have also been added into 
the ZEUS-JETS PDF fit and this provides an improved determination of the sea 
and gluon PDFs at small-$x$. The new fit including all these new data 
is called the ZEUS09 PDF fit.

\section{Results}
\label{sec:results}
\begin{figure}
\vspace{-1.5cm} 
\centerline{
\epsfig{figure=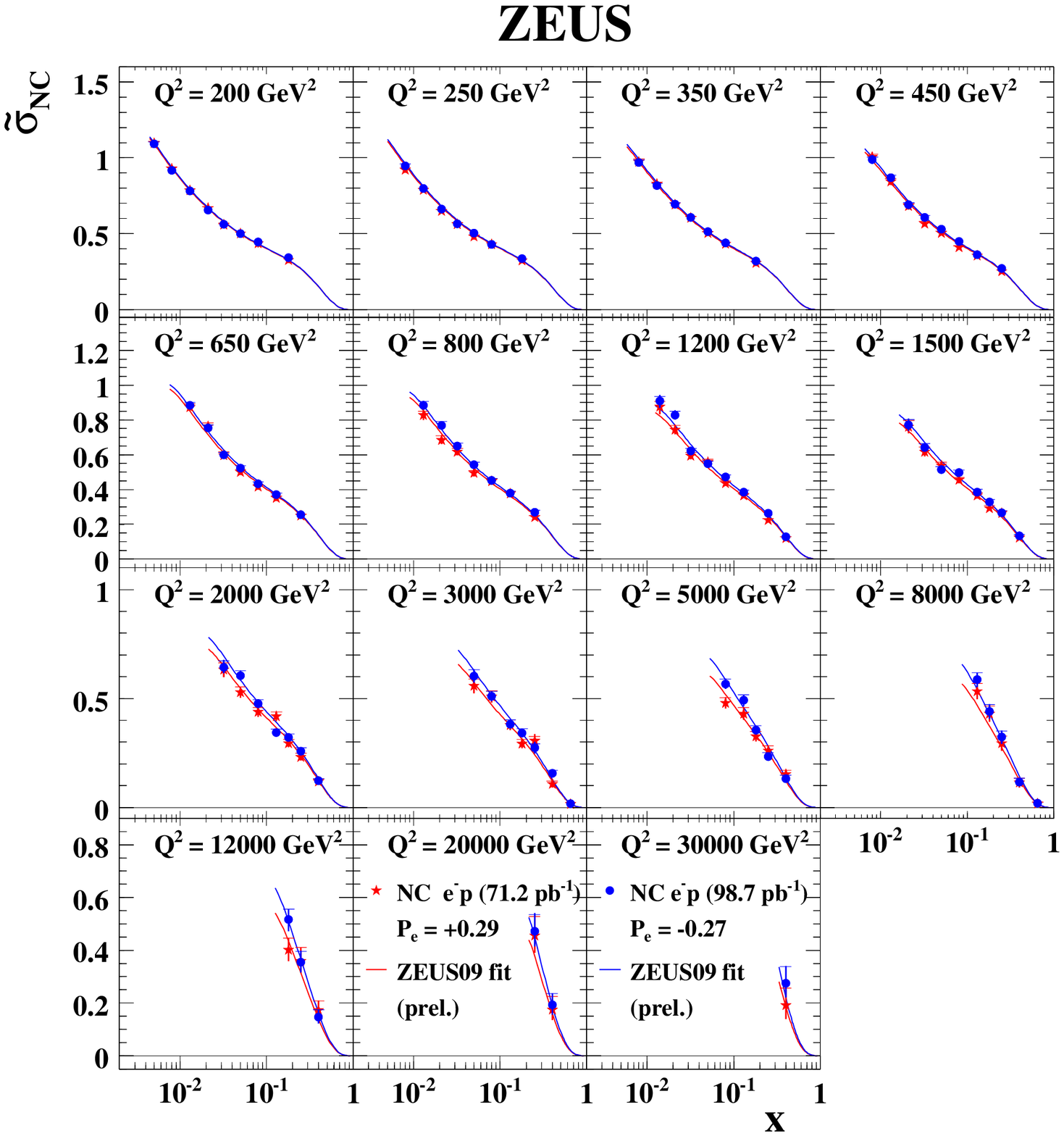,width=0.5\textwidth,height=6.5cm}
\epsfig{figure=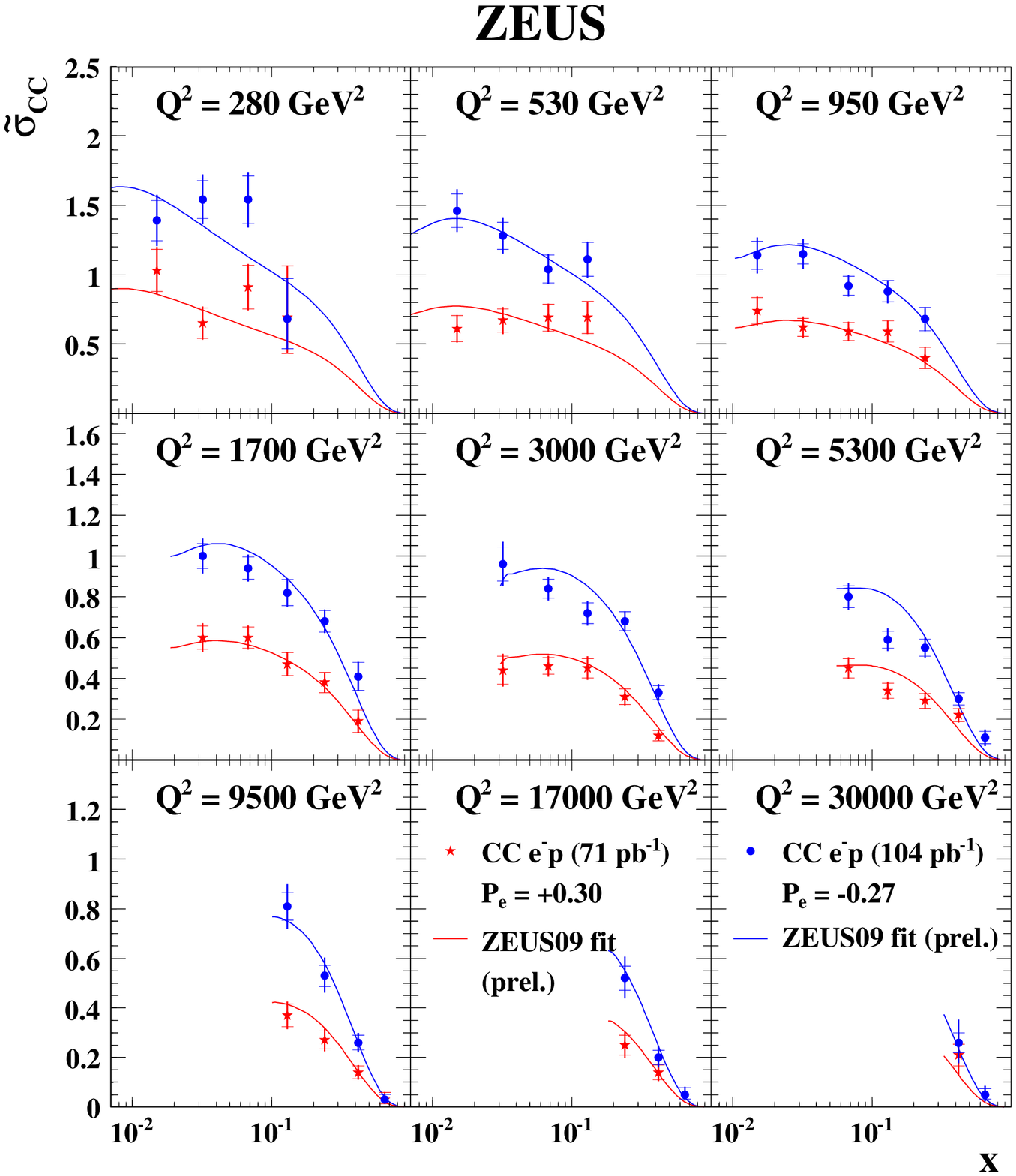,width=0.5\textwidth,height=6.5cm}}
\caption {ZEUS NC (left-hand side) and CC (right-hand side)
 $e^-p$ data from HERA-II running with polarised beams. 
The predictions of the ZEUS09 fit are superimposed}
\label{fig:zeuspol}
\end{figure}

\begin{figure}
\vspace{-1.5cm} 
\centerline{
\epsfig{figure=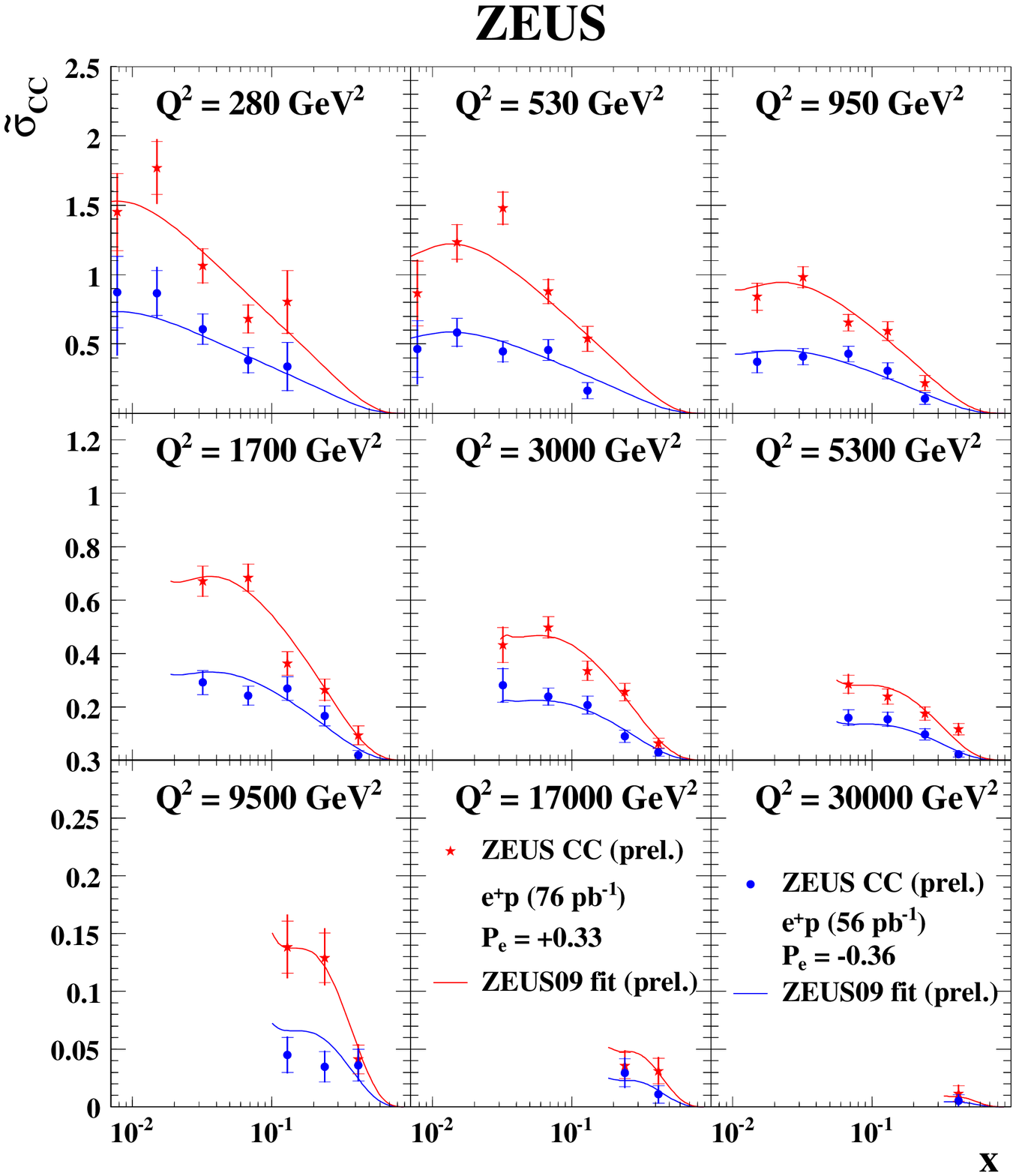,width=0.5\textwidth,height=6.5cm}
\epsfig{figure=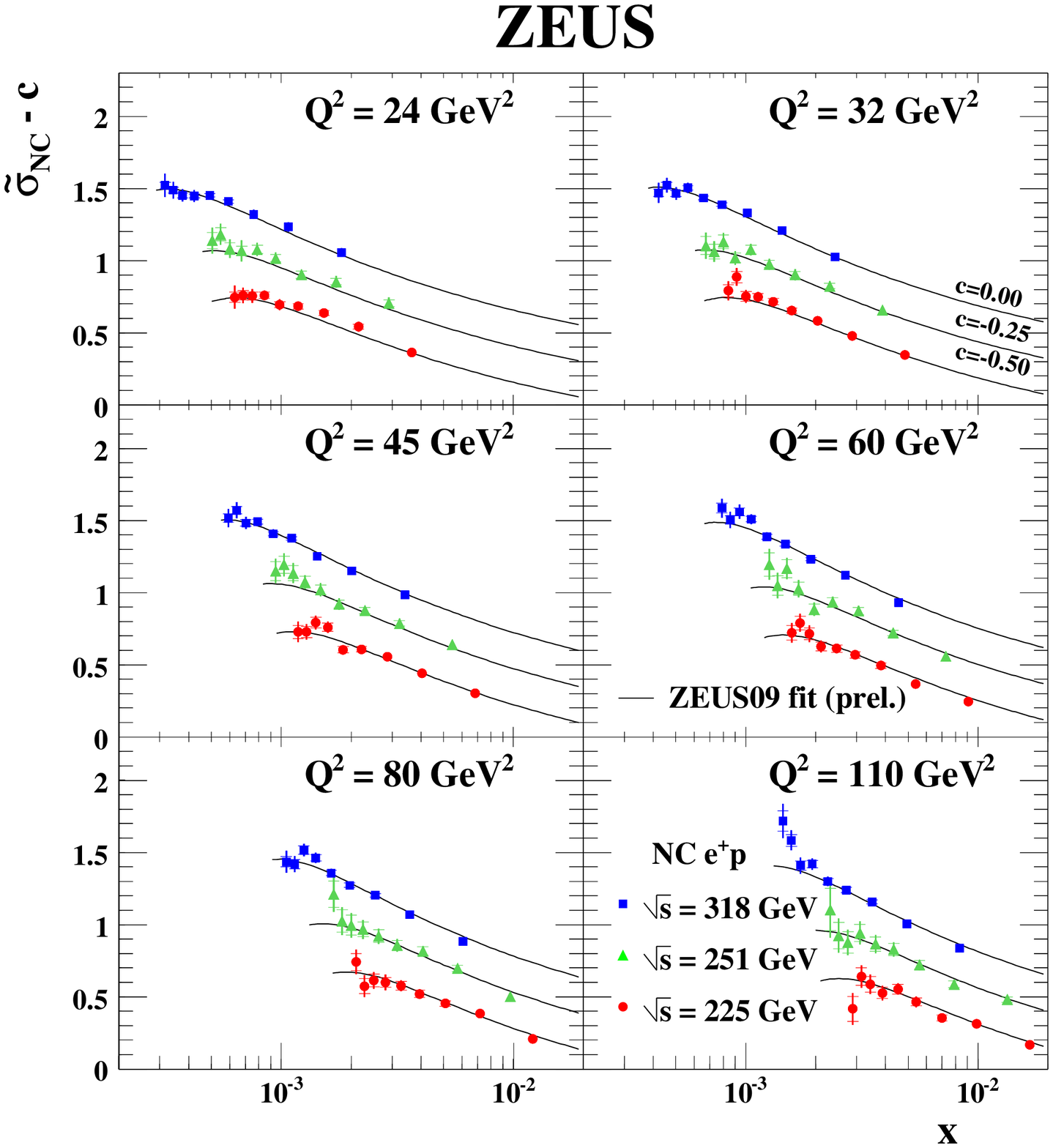,width=0.5\textwidth,height=6.5cm}}
\caption {ZEUS CC $e^+p$ data from HERA-II running with polarised beams 
(left-hand side). ZEUS NC $e^+p$ data from 2007 HERA-II running with three 
different proton beam energies (right-hand side). 
The predictions of the ZEUS09 fit are superimposed}
\label{fig:zeuspolccep}
\end{figure}

Fig~\ref{fig:zeuspol} (left-hand side) shows ZEUS data on NC $e^-p$ double differential 
cross-sections from the 2005-2006 running period.  
There are $99 pb^{-1}$ of negatively polarised data ($P_e=-0.27$)
 and $71 pb^{-1}$ of positively polarised data ($P_e=+0.29$).
This figure (right-hand side) also shows ZEUS data on CC $e^-p$ double differential 
cross-sections from the 2004-2006 running period.  
There are $104 pb^{-1}$ of negatively polarised data ($P_e=-0.27$)
 and $71 pb^{-1}$ of positively polarised data  ($P_e=+0.30$). 
Fig~\ref{fig:zeuspolccep} (left-hand side) shows ZEUS data on CC $e^+p$ double differential 
cross-sections from the 2006-2007 running period.  
There are $56 pb^{-1}$ of negatively polarised data ($P_e=-0.36$)
 and $76 pb^{-1}$ of positively polarised data  ($P_e=+0.33$). 
The same figure (right-hand side) shows ZEUS data on NC $e^+p$ double differential 
cross-sections for three proton beam energies corresponding to,
$\surd{s} = 318, 251, 225~$GeV, from the 2007 running period. There are
$44.5, 7.1, 14.0$pb$^{-1}$ of data for each beam energy respectively.
After the addition of these new data the $577$ data points in the 
ZEUS-JETS PDF fit has increased to $1060$ data points in the new ZEUS09 PDF fit.
The results of this ZEUS09 fit are superimposed on the data 
in all these figures. The $\chi^2$ per degree of freedom of this fit is 0.97.
The agreement of all of the polarised data with the fit is a confirmation not 
only of the validity of the NLO QCD in the DGLAP formalism but also of the 
electroweak predictions of the Standard Model in a space-like process.

\begin{figure}[tbp]
\vspace{-1.5cm} 
\centerline{
\epsfig{figure=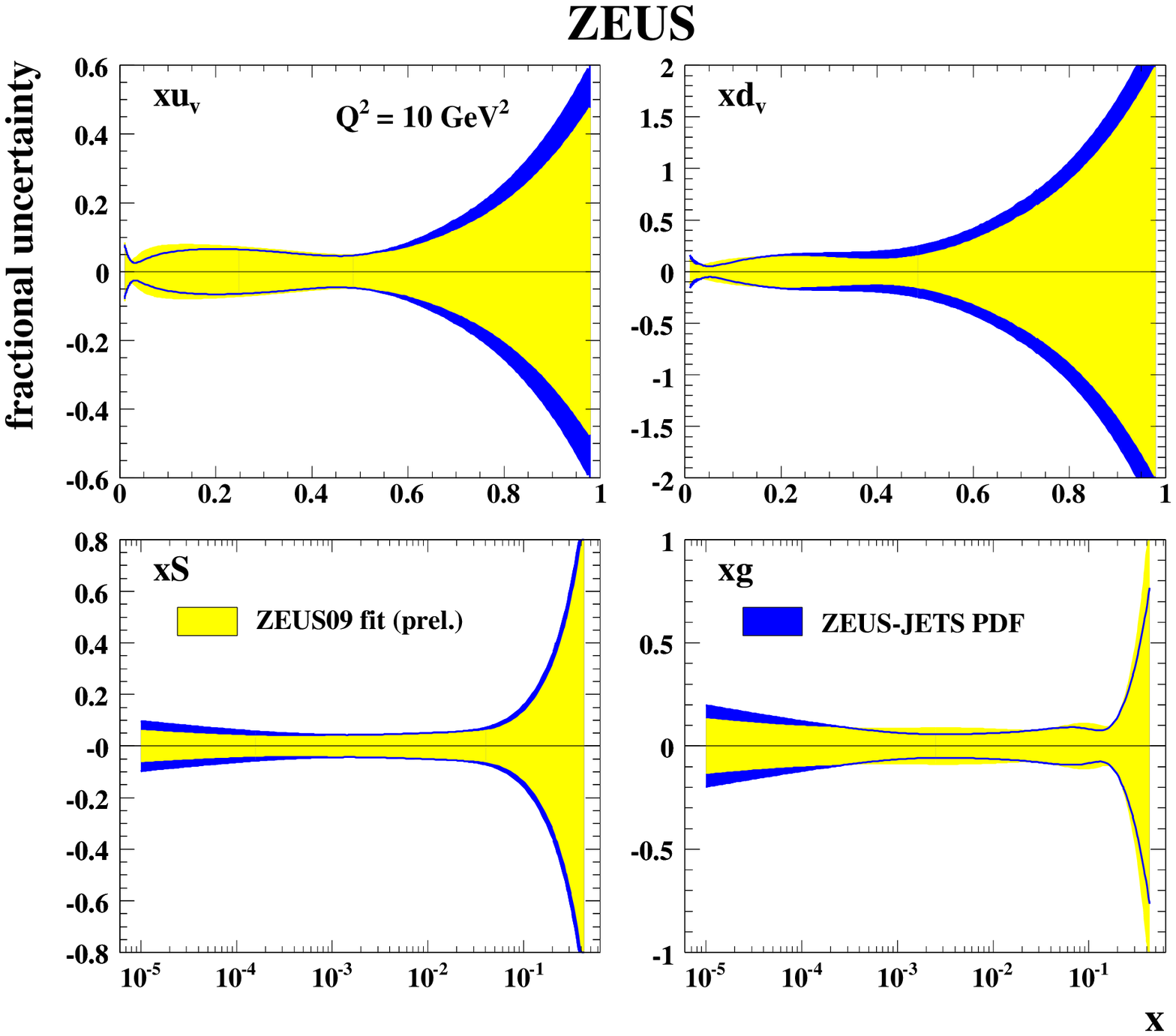,width=0.5\textwidth,height=6.5cm}
\epsfig{figure=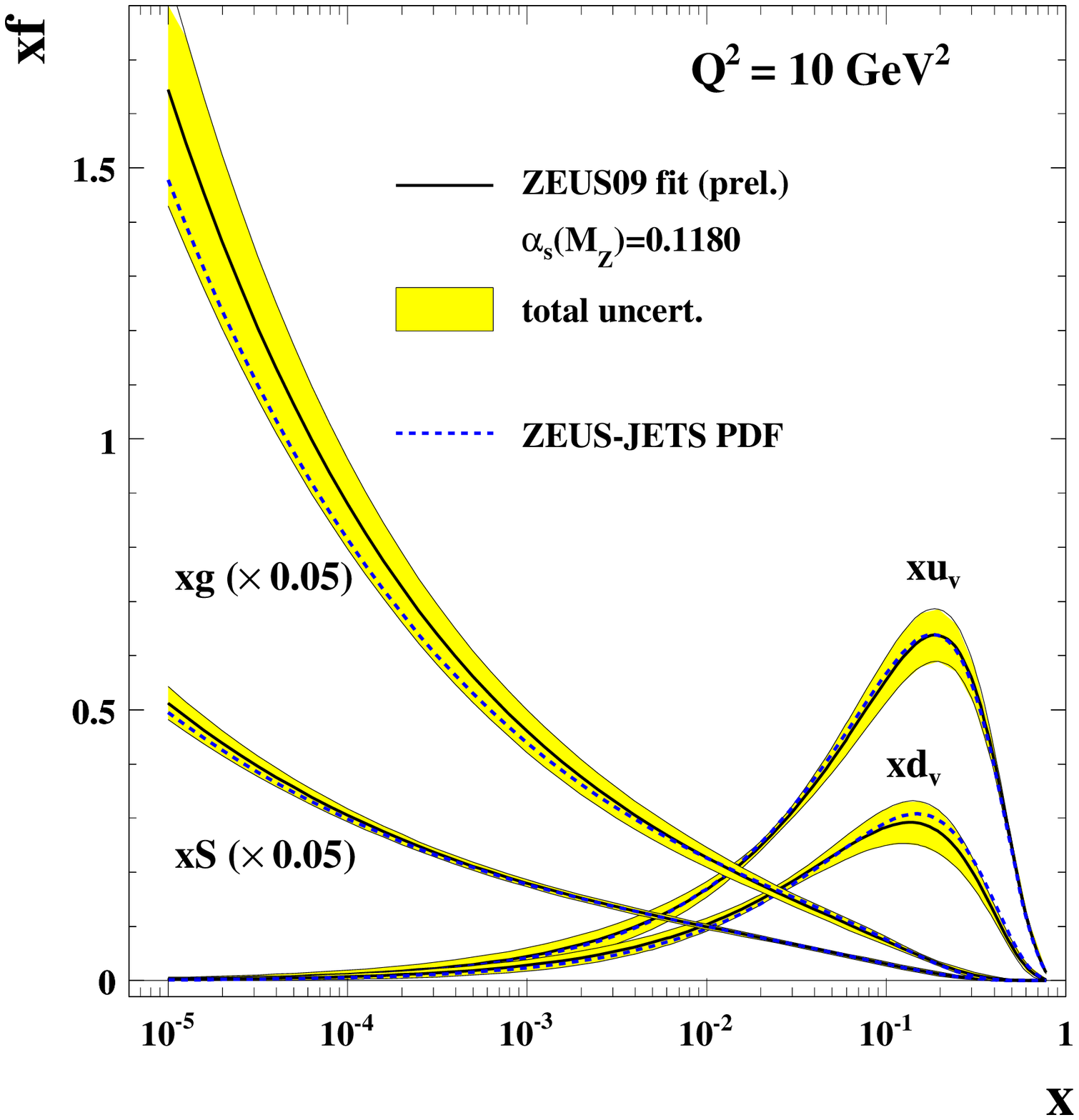,width=0.5\textwidth,height=6.5cm}
}
\caption {Left hand side: The fractional 
uncertainties of the ZEUS09 PDFs 
compared to those of the ZEUS-JETS PDFs. 
Right hand side: The PDFs extracted from the ZEUS09 fit compared to 
the ZEUS-JETS fit at $Q^2=10$GeV$^2$.  }
\label{fig:zpPDF}
\end{figure}

Fig.~\ref{fig:zpPDF} (left-hand-side) illustrates the further impact of adding these new data
 by comparing the fractional uncertainties of the PDFs extracted from 
a fit including these data with those extracted from the ZEUS-JETS PDF. 
The improvement in the $u$-valence quark at high $x$ 
comes from the addition of the 
NC and CC $e^-p$ data from the 2004-2006 running. Both these processes are
 $u$ quark dominated at large $x$. The $d$-valence 
quark uncertainty is also reduced significantly at large $x$. 
This improvement derives from the  CC $e^+p$ data which are 
$d$ quark dominated at large $x$. Finally the improvement in the low-$x$ gluon 
and sea PDFs come from the addition of the NC $e^+p$ 
lower $Q^2$ data run at three different beam energies. This 
final addition of data completes the ZEUS09 PDF fit. The low-$x$ scale of 
the figure is extended to illustrate the improvement 
which is expected from adding these high-$y$ data.  
The PDFs extracted from the ZEUS09 fit are also compared to those of the 
ZEUS-JETS fit in Fig~\ref{fig:zpPDF} (right-hand side). The central 
values of the fit  are very compatible with the 
ZEUS-JETS fit, but the gluon PDF is a little steeper 
indicating the impact of the low-$Q^2$ data on the gluon PDF.

\section{Summary}
\label{sec:summary}
The inclusion of high-$Q^2$ NC and CC $e^-p$ data into the ZEUS-JETS PDF 
fit results in an improved determination of the $u$-valence PDF. The further 
inclusion of high-$Q^2$ CC $e^+p$ data results in an improved determination of 
the $d$-valence PDF. These data were run with polarised lepton beams and 
the cross-section data for the different polarisations provides a spectacular 
confirmation of the Standard Model electroweak predictions in a space-like 
process. Finally the inclusion of lower-$Q^2$ data run at
three different proton beam energies yields an  
improved determination of the low-$x$ sea and gluon PDFs. The new fit 
including all these data is called the ZEUS09 PDF fit~\cite{url}.

\section{Bibliography}
 

\begin{footnotesize}



%

\end{footnotesize}


\end{document}